# Influence of superconducting leads energy gap on electron transport through double quantum dot by Markovian quantum master equation approach

E. Afsaneh and H. Yavari
Department of Physics, University of Isfahan, HezarJarib, Isfahan 81746, Iran

Abstract

The superconducting reservoir effect on the current carrying transport of a double quantum dot in Markovian regime is investigated. For this purpose, a quantum master equation (QME) at finite temperature is derived for the many-body density matrix of an open quantum system. The dynamics and the steady-state properties of the double quantum dot system for arbitrary bias are studied. We will show that how the populations and coherencies of the system states are affected by superconducting leads. The energy parameter of system contains essentially four contributions due to dots system-electrodes coupling, intra dot coupling, two quantum dots inter coupling and superconducting gap. The coupling effect of each energy contribution is applied to currents and coherencies results. In addition, the effect of energy gap is studied by considering the amplitude and lifetime of coherencies to get more current through the system.

Key Words: Quantum Master Equation, Superconducting Gap, Quantum Dot

1. Introduction
Recent years, it has received remarkable progress in experimental techniques to study the electron transport through nanoscopic systems where semiconductor quantum dots (QDs) are connected to superconducting electrodes [1-5].When the coupling in superconductor is sufficient, the Cooper pairs in low temperatures develop global coherency, which suppress the global macroscopic superconductivity. These studies are desirable for very sensitive and controllable coherent switching devices which make it possible to investigate quantum computing and different quantum effects of fundamental physics, such as single electron tunneling, quantum phase transition, and macroscopic condensation [6-9].
Keldysh nonequilibrium Green's functions (NEGF) [10,11], scattering theory(ST) [12,13] and quantum master equation (QME)[14-17]approaches can be used to study the nonequilibrium electron transport through a quantum system (QD or single molecule). ST is limited to elastic processes while the NEGF can treat in elastic and inelastic ones. Both approaches are applied to the tunneling coupling exactly, but usually the correlations inside the dot are fully neglected or the mean field and perturbation theory is used. QME is an alternative tool for studying the irreversible dynamics of quantum systems coupled to a macroscopic environment. This approach treats the correlations inside the dot very accurately (even exactly in the case of model systems) but the tunneling is usually considered in the Born-Markov approximation.
Markovian quantum master equation [18,19] provides an intuitive understanding the non-equilibrium transport problem of the system dynamics [20]and has been used in various fields such as quantum optics[21,22], solid state physics [23],and chemical dynamics [24].For example, electrons tunneling through molecules, coupled QDs [25] and superconducting systems have been studied by this approach.
Recently, QME approach has been applied to investigate electron tunneling through molecules or coupled QDs [26, 27] such as, studying the rectification properties of a system of coupled QDs by analyzing the occupation of two-electron triplet states as a function of the ratio of the interdot coupling [28], deriving a hierarchy of QMEs to study the effects of quantum coherence and Coulomb blockade on steady state electron transport in the high bias limit [9], describing the direct tunneling (where the system never gets charged) in quantum junctions [29], generalizing the standard rate equation beyond the second order perturbation in system-lead coupling [30] and considering the spin polarization of the electrons to study Pauli blockade [28] and magnetotransport [31] in QDs.
Superconducting systems has been also investigated by QME techniques recently, including the proximity effect in one dimensional wires [32], transport properties of a spin degenerate single QD connected to superconducting reservoirs [6] and transport through quantum dots coupled to superconducting leads [33].

The importance of increasing the coherency and reduction of dissipation in quantum transport causes to apply the non-dissipative condensate state of superconductors as reservoirs and using two coupled QDs as system. Double quantum dot (DQD) [34], as artificial atoms is able to confine electrons to form effective two level system due to manipulating and controlling coherent transport [35,36].

In this paper, we present a derivation of the master equation [37] in the case of DQD system with the superconducting electrodes which are described by Bogoliubov-de Gennes Hamiltonians [38]. Thus we deal with four different energy parameters: the tunneling coupling between the whole dots system and electrodes, the strength of electronic correlations inside the dot, the inter coupling between two QDs and the superconducting gap energy according the BCS theory [39, 40] in the electrodes. We study in detail the quasiparticle current of the model and the effect of quantum coherences on electric transport.

The effect of double quantum dot is introduced by inter-dots coupling energy in comparison with degenerate single one [6] which affects the coherency and current. We show that the coherency time evolution and current are expressed by the presence of the energy gap of superconducting leads.

The outline of the paper is as follows. In Sec. 2, we discuss the quantum master equation approach. In Sec. 3, we present our model Hamiltonian and calculate the current without coherency and with considering coherency analytically. In latter case, the importance of interchange energy coupling between two dots is shown well by introducing general and approximate forms. The dynamics of population and coherencies are also performed. We present the numerical results in section 4, discuss the effect of quantum coherences on the current by comparing various mentioned forms for both normal and superconducting leads. Then, we investigate the energy gap effect on coherency oscillation. Finally, conclusions are offered in Sec. 5.

2. Quantum master equation for a double quantum dot connected to superconducting leads

We consider a quantum mechanical system S weakly coupled to a reservoir B. The Hamiltonian of the total system is given by

$$H = H_S + H_B + H_I, \qquad (1)$$

where $H_S$ and $H_B$ are the free Hamiltonian of the system and the reservoir respectively. $H_I$ describes the interaction between the system and the reservoir, which can be written generally as

$$H_I = \sum_\alpha A_\alpha \otimes B_\alpha, \qquad (2)$$

here $A_\alpha$ and $B_\alpha$ are the operators that act on the system and bath respectively and commute $[A_\alpha, B_\alpha] = 0$. The Hermiticity of $H_I$ is expressed by the conditions $A_\alpha^\dagger = A_{\alpha'}$ and $B_\alpha^\dagger = B_{\alpha'}$. The interaction Hamiltonian $H_I$ decomposes into eigen operators of the system Hamiltonian $H_S$. The eigenvalues of $H_S$ is represented by $\epsilon$ and the projection onto the eigen space belonging to the eigenvalue $\epsilon$ is denoted by $|\epsilon\rangle\langle\epsilon|$. Thus, $A_\alpha(\omega) = \sum_{\epsilon'-\epsilon=\omega} |\epsilon\rangle\langle\epsilon|A_\alpha|\epsilon'\rangle\langle\epsilon'|$, where the sum is extended over all energy eigenvalues $\epsilon'$ and $\epsilon$ of $H_S$ (Appendix A).

To analyses the dynamical behavior of the populations and coherences of system, it is convenient to formulate the density matrix dynamics of the total system in the interaction picture which is governed by the von Neumann equation and used the reduced density matrix $\rho_s(t) = tr_B[\rho(t)]$. $\rho(t)$ denotes the density matrix of the composite system and $tr_B$ is the partial trace taken over the bath(environment).

To derive the master equation of an open quantum system, we use the Born-Markov approximation. The Born approximation assumes the coupling between the system and the reservoir is weak, such that the influence of the system on the reservoir is small (weak-coupling approximation). Thus, we have $\rho(t) \approx \rho_s(t) \otimes \rho_B$, in which the density matrix of the reservoir $\rho_B$ is only negligibly affected by the interaction.

The Markovian approximation provides a description on a coarse-grained time scale (local in time).The relevant physical condition for the Born-Markov approximation is that the bath correlation time is small compared to the relaxation time of the system. Then the quantum master equation in the interaction picture can be written as [30](Appendix A)

$$\frac{d\rho_s(t)}{dt} = -i[H_{LS}, \rho_s(t)] + D\rho_s(t), \tag{3}$$

where the Hermitian operator

$$H_{LS} = \sum_{\omega}\sum_{\alpha,\beta} S_{\alpha\beta}(\omega) A_\alpha^\dagger(-\omega) A_\beta(\omega), \tag{4}$$

is the dynamics part of Hamiltonian. This term is often called the Lamb shift Hamiltonian since it leads to a Lamb-type renormalization of the unperturbed energy levels induced by the system-reservoir coupling. Note that the Lamb shift Hamiltonian commutes with the unperturbed system Hamiltonian, $[H_S, H_{LS}] = 0$. The dissipator of the master equation takes the form

$$D\rho_s(t) = \sum_{\omega}\sum_{\alpha,\beta} \gamma_{\alpha\beta}(\omega)\left[2A_\beta(\omega)\rho_s(t)A_\alpha^\dagger(-\omega) - \{A_\alpha^\dagger(-\omega)A_\beta(\omega), \rho_s(t)\}\right]. \tag{5}$$

We define the bath correlation function $\Gamma_{\alpha\beta}(\omega)$ as

$$\Gamma_{\alpha\beta}(\omega) = \int_0^\infty e^{i\omega s} ds \langle B_\alpha^\dagger(s) B_\beta(0) \rangle = \gamma_{\alpha\beta}(\omega) + i S_{\alpha\beta}(\omega), \tag{6}$$

From Eq. (6) we can write

$$\gamma_{\alpha\beta}(\omega) = \frac{1}{2}(\Gamma_{\alpha\beta}(\omega) + \Gamma_{\beta\alpha}^*(\omega)) = \int_{-\infty}^{+\infty} e^{i\omega s} ds \langle B_\alpha^\dagger(s) B_\beta(0) \rangle, \tag{7}$$

$$S_{\alpha\beta}(\omega) = \frac{1}{2i}(\Gamma_{\alpha\beta}(\omega) - \Gamma_{\beta\alpha}^*(\omega)), \tag{8}$$

Here we use $\int_0^\infty ds e^{-i\varepsilon s} = \pi\delta(\varepsilon) + iP\frac{1}{\varepsilon}$, where $P$ denotes the Cauchy principal value. We note that the master equation (3), can be brought into Lindblad form by diagonalization of the matrices $\gamma_{\alpha\beta}(\omega)$ defined in Eq. (9) and use the rotating wave approximation (RWA)[39], which preserve the positivity and Hermiticity of the density matrix.

3. Model calculations

We consider a double quantum dot (Fig.1) with $H_S$,

$$H_S = \sum_s \varepsilon_s c_s^\dagger c_s, \tag{9}$$

connected two uncorrelated one-dimensional superconducting leads described by the Bogoliubov-de Gennes Hamiltonian $H_B$

$$H_B = \sum_k \varepsilon_k (b_{k\uparrow}^\dagger b_{k\uparrow} - b_{-k\downarrow}^\dagger b_{-k\downarrow}) + \Delta(b_{-k\downarrow} b_{k\uparrow} + b_{k\uparrow}^\dagger b_{-k\downarrow}^\dagger), \tag{10}$$

here $b_{k,\sigma}^\dagger$ ($b_{k,\sigma}$) are creation(annihilation) operator for an electron with spin $\sigma = \uparrow, \downarrow$, single-particle energy $\varepsilon_k$ and $\Delta$ the gap energy (order parameter), which governs the superconducting properties of the leads. The index $k$ runs over the modes of the left and right leads and $\Delta$ may have different values for the left and right leads. By using the Bogoliubov transformation

$$b_{k\uparrow} = u(\varepsilon_k) a_{k\uparrow} - v(\varepsilon_k) a_{-k\downarrow}^\dagger, \quad b_{-k\downarrow}^\dagger = v^*(\varepsilon_k) a_{k\uparrow} + u^*(\varepsilon_k) a_{-k\downarrow}^\dagger, \tag{11}$$

$$|u(\varepsilon_k)|^2 = \frac{1}{2}\left(1 + \frac{\varepsilon_k}{\omega(\varepsilon_k)}\right), |v(\varepsilon_k)|^2 = \frac{1}{2}\left(1 - \frac{\varepsilon_k}{\omega(\varepsilon_k)}\right), \tag{12}$$

$$\omega^2(\varepsilon_k) = \varepsilon_k^2 + \Delta^2. \tag{13}$$

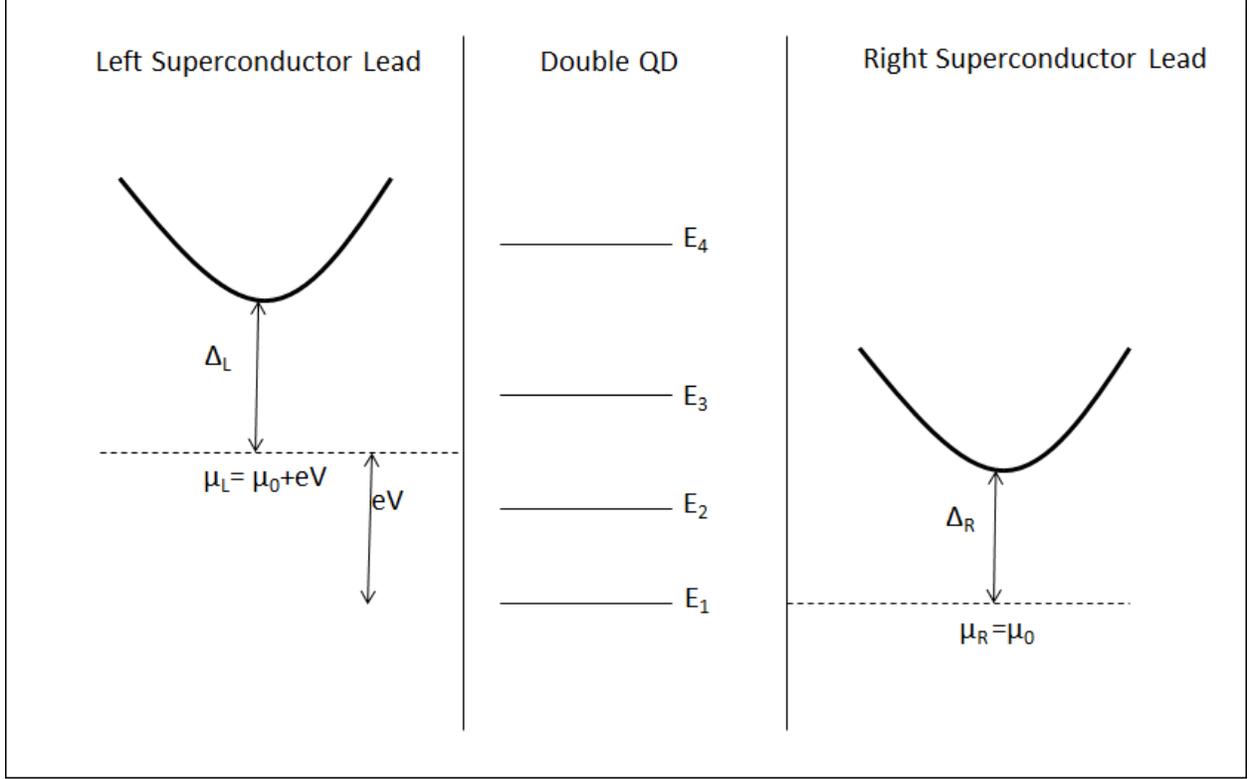

Fig.1. Lead-system-lead configuration. $\mu_L$ and $\mu_R$ are the chemical potentials of the left and right leads. $E_1$, $E_2$, $E_3$ and $E_4$ are the energies of the system many-body states.

Eq.(10) can be written as

$$H_B = \sum_\delta \sum_{k\sigma} \omega(\varepsilon_{k\sigma,\delta}) a^\dagger_{k\sigma,\delta} a_{k\sigma,\delta}, \tag{14}$$

where $\delta = L, R$. Changing the summation over k to integral

$$\frac{1}{V}\sum_k \to \int \frac{d^3k}{(2\pi)^3} = \int N(\varepsilon_k) d\varepsilon_k = \int \rho(\omega) d\omega, \tag{15}$$

where $\rho(\omega)$ is the density state excitation for both left and right leads. According to BCS theory, the density of states is

$$\rho_\nu(\omega_\nu) = \left(\frac{d\omega_\nu}{d\varepsilon_\nu}\right)^{-1} = \Theta(|\omega_\nu| - \Delta_\nu)\frac{|\omega_\nu|}{\sqrt{\omega_\nu^2 - \Delta_\nu^2}}; \nu = L, R, \tag{16}$$

$\Theta(\omega)$ is the Heaviside step function in superconducting systems (for normal metal systems $\Delta = 0$)

Now we introduce two coupled QDs system connected the left and right leads with the chemical potentials $\mu_L$ and $\mu_R$, respectively. The system Hamiltonian is

$$H_S = \sum_{ij=1,2} \omega_{ij} c_i^\dagger c_j = \omega_{11} c_1^\dagger c_1 + \omega_{12} c_1^\dagger c_2 + \omega_{21} c_2^\dagger c_1 + \omega_{22} c_2^\dagger c_2, \qquad (17)$$

Where $\omega_{11}$ and $\omega_{22}$ are system orbital energies of each QDs and $\omega_{12}$ and $\omega_{21}$, the interdot energies, are dependent on localization (weak coupling) or delocalization (strong coupling) of orbital system.

By ignoring the charging effects due to electron-electron interactions (Coulomb-blockade) [41] in the system, the interaction Hamiltonian $H_I$ becomes

$$H_I = \sum_\nu \sum_{k\sigma} [T_{k\sigma,\nu} c_s^\dagger a_{k\sigma,\nu} + T_{k\sigma,\nu}^* a_{k\sigma,\nu}^\dagger c_s]; \nu = L, R; s = 1,2, \qquad (18)$$

where $T_{k\sigma,\nu}$ is the tunneling coupling between the dots system and electrodes. From equation (3) the projector operators of the DQD system is

$$A(\omega_1, \omega_2) = \sum_{\omega_1' - \omega_1 = \omega_1} \sum_{\omega_2' - \omega_2 = \omega_2} |\omega_1, \omega_2\rangle\langle \omega_1, \omega_2 | A_1 A_2 | \omega_1', \omega_2'\rangle\langle \omega_1', \omega_2' |. \qquad (19)$$

By using Eq. (19) into Eq. (5), the dissipator operator becomes

$$D\rho_s(t) = \sum_\omega \sum_{i,j=1,2} \gamma_{ij}(\omega_i, \omega_j) \left[ 2A_j(\omega_j)\rho_s(t) A_i^\dagger(-\omega_i) - \{A_i^\dagger(-\omega_i) A_j(\omega_j), \rho_s(t)\} \right]. \qquad (20)$$

From Eq. (4), the Lamb shift part can be written as

$$H_{LS} = \sum_{i,j=1}^{2} \sum_\omega S_{ij}(\omega) A_i^\dagger(-\omega_i) A_j(\omega_j). \qquad (21)$$

The nonzero projections of Eq. (19) become

$$A_1(\omega_1, 0) = |0,0\rangle\langle 1,0|,\ A_1(-\omega_1, 0) = |1,0\rangle\langle 0,0|,\ A_1(\omega_1, \omega_2) = |0,1\rangle\langle 1,1|,\ A_1(-\omega_1, \omega_2) = |1,1\rangle\langle 0,1|,$$
$$A_2(0, \omega_2) = |0,0\rangle\langle 0,1|,\ A_2(0, -\omega_2) = |0,1\rangle\langle 0,0|,\ A_2(\omega_1, \omega_2) = |1,0\rangle\langle 1,1|,\ A_2(\omega_1, -\omega_2) = |1,1\rangle\langle 1,0|. \qquad (22)$$

By considering the positive and negative parts of $\omega$ and using the projections Eq. (22) and their complex conjugate, the dissipator operator and Lamb shift parts, respectively Eqs. (20, 21) are obtained. The full system density matrix has 16 components but in the reduced space, it becomes a six components density matrix. Thus, the quantum master equation, Eq. (3), for reduced density matrix is given by

$$\dot{\rho}(t) = \hat{M}\rho(t). \qquad (23)$$

Where the density matrix in the reduced space is $\rho = (\rho_{00,00}, \rho_{01,01}, \rho_{10,10}, \rho_{11,11}, \rho_{01,10}, \rho_{10,01})$ ($\hat{M}$ matrix elements are considered in Appendix B).

At the steady state ($\dot{\rho} = 0$), by using $\rho_{00,00} + \rho_{01,01} + \rho_{10,10} + \rho_{11,11} = 1$ we have $\rho_{00,00} = \gamma_{11}^R(\omega_1, \omega_1)\gamma_{22}^R(\omega_2, \omega_2)/d$, $\rho_{01,01} = \gamma_{11}^R(\omega_1, \omega_1)\gamma_{22}^L(-\omega_2, -\omega_2)/d$, $\rho_{10,10} = \gamma_{11}^L(-\omega_1, -\omega_1)\gamma_{22}^R(\omega_2, \omega_2)/d$ and $\rho_{11,11} = \gamma_{11}^L(-\omega_1, -\omega_1)\gamma_{22}^L(-\omega_2, -\omega_2)/d$, where $d = [\gamma_{11}^L(-\omega_1, -\omega_1) + \gamma_{11}^R(\omega_1, \omega_1)][\gamma_{22}^L(-\omega_2, -\omega_2) + \gamma_{22}^R(\omega_2, \omega_2)]$. Here, $\gamma_{ss'}^\nu(\omega_s, \omega_{s'})$ is

$$\gamma_{ss'}^\nu(\omega_s, \omega_{s'}) = \eta_{\nu,ss'} \rho_\nu(\omega_{ss'})(1 - f_\nu(\omega_{ss'})),$$
$$\gamma_{ss'}^\nu(-\omega_s, -\omega_{s'}) = \eta_{\nu,ss'} \rho_\nu(\omega_{ss'}) f_\nu(\omega_{ss'}), \qquad (24)$$

in which $\eta_{\nu,ss'}$ is the coefficient coupling, $f(\omega_k) = 1/(1 + e^{\omega(\omega_k)/k_B T})$ is the distribution function of bath, $\nu = L, R$ denotes the left and right leads and $s, s' = 1, 2$ refers to quantum dot 1 and 2. The formalism of the currents may be written as

$$I_X(t) = e\langle N|\widehat{M}_X|\rho(t)\rangle, \tag{25}$$

here $\widehat{M}_X$ is the contribution from lead $X$ to the matrix $\widehat{M}$ such that $\widehat{M} = \widehat{M}_L + \widehat{M}_R$. Since $\gamma(\dot{\omega})$ ($\gamma(-\dot{\omega})$) is related to the outflux (influx) of electrons from (to) the system, we can separate the current as $I_X = I_X^{in} + I_X^{out}$

$$I_X^{in} = e\langle N|\widehat{M}_X^{in}[\gamma^L(-\dot{\omega})]|\rho(t)\rangle, \tag{26}$$

$$I_X^{out} = e\langle N|\widehat{M}_X^{out}[\gamma^L(\dot{\omega})]|\rho(t)\rangle, \tag{27}$$

where $\widehat{M}_X[\gamma(\dot{\omega})]$ and $\widehat{M}_X[\gamma(-\dot{\omega})]$ include $\gamma^X(\dot{\omega})$ and $\gamma^X(-\dot{\omega})$, respectively.

By considering all elements of $\widehat{M}$ matrix, from Eqs.(26) and (27), the general form of the currents with coherency can be written as

$$I_{L,general}^{in} = 2\frac{\gamma_{11}^L(-\dot{\omega}_1,-\dot{\omega}_1)\gamma_{11}^R(\dot{\omega}_1,\dot{\omega}_1)}{\gamma_{11}^L(-\dot{\omega}_1,-\dot{\omega}_1)+\gamma_{11}^R(\dot{\omega}_1,\dot{\omega}_1)} + 2\frac{\gamma_{22}^L(-\dot{\omega}_2,-\dot{\omega}_2)\gamma_{22}^R(\dot{\omega}_2,\dot{\omega}_2)}{\gamma_{22}^L(-\dot{\omega}_2,-\dot{\omega}_2)+\gamma_{22}^R(\dot{\omega}_2,\dot{\omega}_2)} + 2\gamma_{12}^L(-\dot{\omega}_1,-\dot{\omega}_2)\rho_{01,10} + 2\gamma_{21}^L(-\dot{\omega}_2,-\dot{\omega}_1)\rho_{01,10}^* \tag{28}$$

$$I_{L,general}^{out} = -2\frac{\gamma_{11}^L(-\dot{\omega}_1,-\dot{\omega}_1)\gamma_{11}^L(\dot{\omega}_1,\dot{\omega}_1)}{\gamma_{11}^L(-\dot{\omega}_1,-\dot{\omega}_1)+\gamma_{11}^R(\dot{\omega}_1,\dot{\omega}_1)} - 2\frac{\gamma_{22}^L(-\dot{\omega}_2,-\dot{\omega}_2)\gamma_{22}^L(\dot{\omega}_2,\dot{\omega}_2)}{\gamma_{22}^L(-\dot{\omega}_2,-\dot{\omega}_2)+\gamma_{22}^R(\dot{\omega}_2,\dot{\omega}_2)} - 2\gamma_{12}^L(\dot{\omega}_1,\dot{\omega}_2)\rho_{01,10} - 2\gamma_{21}^L(\dot{\omega}_2,\dot{\omega}_1)\rho_{01,10}^*, \tag{29}$$

here $\rho_{01,10,general}$, the general form of coherency is

$$\rho_{01,10,g} = \frac{c-d-ie+f}{ab}, \tag{30}$$

where $a = [\gamma_{11}^L(-\dot{\omega}_1,-\dot{\omega}_1) + \gamma_{11}^R(\dot{\omega}_1,\dot{\omega}_1) + \gamma_{22}^L(-\dot{\omega}_2,-\dot{\omega}_2) + \gamma_{22}^R(\dot{\omega}_2,\dot{\omega}_2) + i(\dot{\omega}_1 - \dot{\omega}_2)]$,
$b = [\gamma_{11}^L(-\dot{\omega}_1,-\dot{\omega}_1) + \gamma_{11}^R(\dot{\omega}_1,\dot{\omega}_1)][\gamma_{22}^L(-\dot{\omega}_2,-\dot{\omega}_2) + \gamma_{22}^R(\dot{\omega}_2,\dot{\omega}_2)]$, $c = 2\gamma_{21}^L(-\dot{\omega}_2,-\dot{\omega}_1)\gamma_{11}^R(\dot{\omega}_1,\dot{\omega}_1)\gamma_{22}^R(\dot{\omega}_2,\dot{\omega}_2)$,
$d = [\gamma_{12}^L(-\dot{\omega}_1,-\dot{\omega}_2) + \gamma_{21}^R(\dot{\omega}_2,\dot{\omega}_1)][\gamma_{11}^R(\dot{\omega}_1,\dot{\omega}_1)\gamma_{22}^L(-\dot{\omega}_2,-\dot{\omega}_2) + \gamma_{11}^L(-\dot{\omega}_1,-\dot{\omega}_1)\gamma_{22}^R(\dot{\omega}_2,\dot{\omega}_2)]$,
$e = (\dot{\omega}_2 - \dot{\omega}_1)[\gamma_{11}^R(\dot{\omega}_1,\dot{\omega}_1)\gamma_{22}^L(-\dot{\omega}_2,-\dot{\omega}_2) - \gamma_{11}^L(-\dot{\omega}_1,-\dot{\omega}_1)\gamma_{22}^R(\dot{\omega}_2,\dot{\omega}_2)]$ and $f = \gamma_{12}^R(\dot{\omega}_1,\dot{\omega}_2)\gamma_{11}^L(-\dot{\omega}_1,-\dot{\omega}_1)\gamma_{22}^L(-\dot{\omega}_2,-\dot{\omega}_2)$.

By using $\gamma_{12}^L(-\dot{\omega}_1,-\dot{\omega}_2) = \gamma_{21}^L(-\dot{\omega}_2,-\dot{\omega}_1)$ and $\gamma_{12}^R(\dot{\omega}_1,\dot{\omega}_2) = \gamma_{21}^R(\dot{\omega}_2,\dot{\omega}_1)$ (due to the same interchange of inter-dots coupling energy, $\dot{\omega}_{21} = \dot{\omega}_{12}$) the approximation form of the currents and coherency can also be obtained from Eqs.(27), (28) and (29).

At steady state ($t \to \infty$), the currents from left and right leads must be equal and opposite in sign ($I_s = I_L = -I_R$). Thus the steady state currents without coherency $I_L^{in}$ and $I_L^{out}$ are obtained analytically from Eqs. (26) and (27), which shows the decoupling of non-interacting QDs like a degenerate single quantum dot currents [6].

The dynamics of population and coherencies are determined by solving the Eq. (22) and diagonalizing the matrix $\widehat{M}$.

$$|\rho(t)\rangle = e^{\xi t}|\rho(0)\rangle = \sum_i g_i e^{\varepsilon_i t}|i\rangle \tag{31}$$

where $\varepsilon_i$ ($|i>$) are the eigenvalues (eigenvectors) of $\hat{M}$ and $g_i = <i|\rho(0)>$. In the RWA, the off-diagonal coherency terms are neglected and then the population dynamics become independent of the coherences which can be obtained analytically (Appendix C).

4. Numerical results

We find the $\hat{M}$ matrix by quantum master equation under the system and leads physical quantities. The dynamics and steady state of populations and coherences are obtained by solving the Eq. (22). We choose the numerical values of the double quantum dot energies, $\varepsilon_1 = 5 meV$ and $\varepsilon_2 = 2 meV$, and temperature $k_B T = 0.2 meV$.

Fig. 2 indicates the effect of change bias on the system states occupation for identical left and right couplings ($\eta_{Ls} = \eta_{Rs}$). We suppose that the chemical potential for the right lead is fixed at the Fermi energy $\mu_0$ while the left one is increased by the bias. For $\mu_0 \leq 0$ and zero bias, there are no electrons in the system. As the bias becomes on, electrons start to move from left to right lead through the system. In low bias ($eV < \varepsilon_2 + \Delta$) all many-body states are unoccupied and the probability of the state $|00\rangle$ is one. By increasing the bias, the states are filling by electrons. For $eV < \varepsilon_1$ only states $|00\rangle$ and $|01\rangle$ are populated. In high bias ($eV < \varepsilon_1 + \Delta$), all system states have the same occupation. For normal metal leads the energy threshold of moving electron is confined in orbital energy states. The filling populated states appear in step-wise. On the other hand, the electrons of superconducting leads carry energies more than the sum of energy state and energy gap, therefore their population states seem the exact step function.

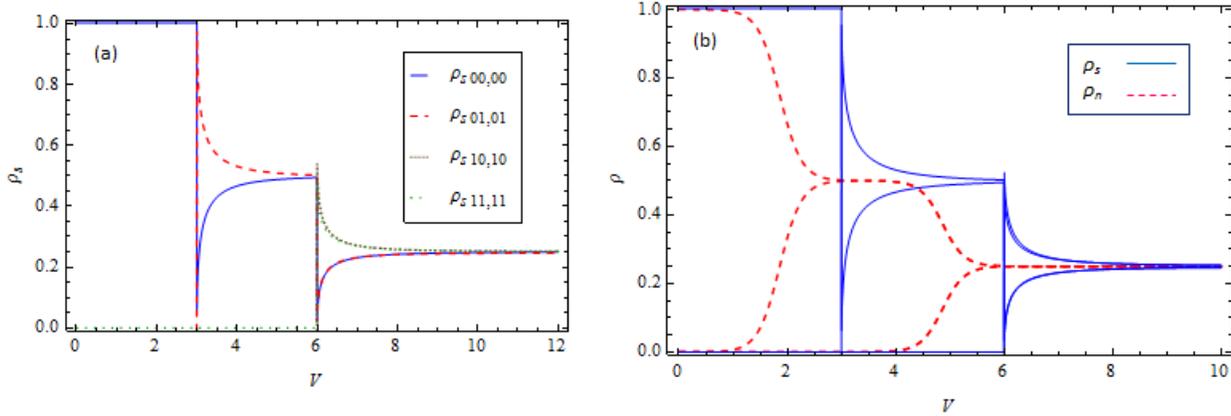

Fig.2. Steady state populations for $\mu_0 = 0$. (a) Superconducting leads. (b) Dashed line corresponds to normal leads and full line corresponds to superconducting leads. The left and right coupling are the same, $\eta_{L1} = \eta_{R1} = 0.4$, $\eta_{L2} = \eta_{R2} = 0.6$ and for superconducting leads $\Delta_{L,R} = 1$. All parameters are in meV.

The steady state current without coherency consideration for both normal metal and superconducting leads are depicted in Fig. 3, such as total current $I_{sL} = I_{inL} + I_{outL}$, input and output current ($I_{inL}$ and $I_{outL}$). The $I_{outL}$ is significant only at resonant energies $eV + \mu_0 = \varepsilon_s$. It can be seen that the current through system to superconducting electrodes at $\varepsilon_s + \Delta$ peaks, represents the proximity effect. Also the current curves show that in high bias where different many-body states are populated similarly, the current for both normal and superconducting leads behave smoothly by increasing bias. It can be expressed that in this regime the intra and inter quantum dot coupling contributions comes to be disappeared.

The effect of the coherences in currents is obtained by solving the QME Eq. (22) and diagonalizing the full 6 × 6 $\hat{M}$ matrix, which we present in the general and approximate formula. In general form, the interchange coupling coefficients of both QDs are different ($\grave{o}_{\nu 12} = \grave{o}_{\nu 21}, \eta_{\nu 12} = \eta_{\nu 21}$), but in approximate one these are equal.

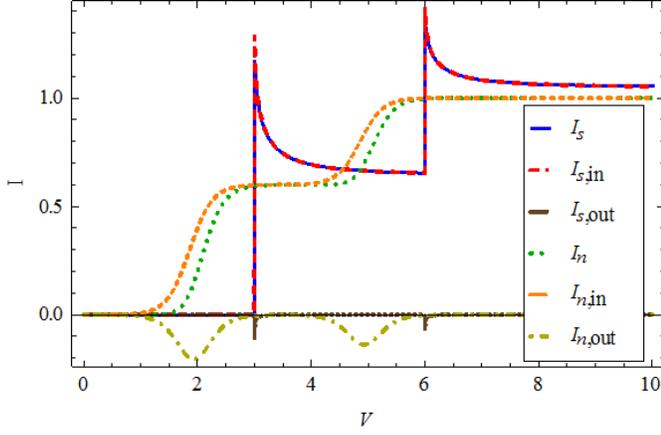

Fig.3. Current without coherency. The left and right coupling are the same, $\eta_{L1} = \eta_{R1} = 0.4$ and $\eta_{L1} = \eta_{R1} = 0.6$ and for superconductor leads $\Delta_{L,R} = 1 meV$ (Current is in units of $me^2V/\hbar$).

Fig. 4 shows the coherency effect in currents for normal leads. Part (a) pointes out the total, input and output currents due to coherency in general form. The currents without coherency and with coherency in both approximate and general formula are compared in part (b). The real and imaginary parts of coherency in general form are shown in (c). Part (d) expresses the coherencies variance due to approximate and general formula which shows positive real part according to the general formula.

It should be mentioned that due to coherences, the backward current $I_{out}$ (dotted line) does not vanish for $eV + \mu_0 = \grave{o}_s$ and it is positive. Although, it is still maximum and negative at the resonances which affect the total current increasingly.

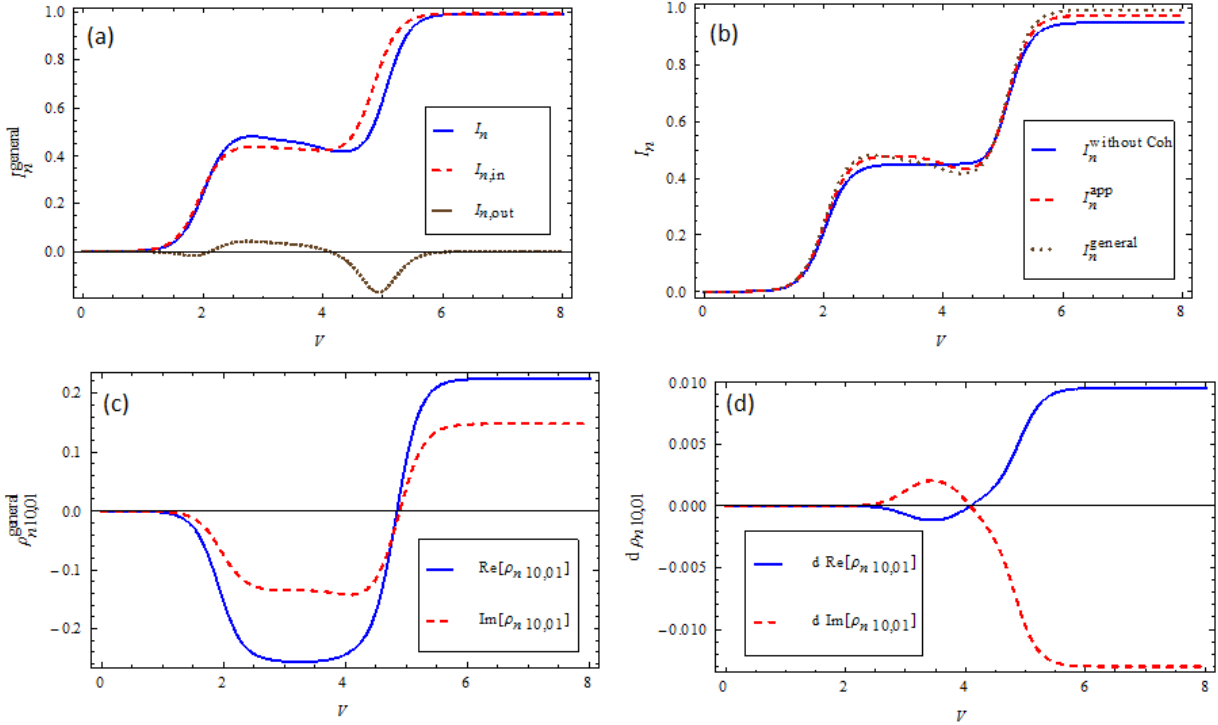

Fig.4. Steady state currents and coherencies of normal leads. (a) Steady state currents with coherences in general formula Eqs. (27) and (28). (b) The comparison with the steady state currents without coherences, approximate formula and general formula. (c) The coherences in general formula Eq. (29). (d) change in the coherences due to approximation and general formula. The couplings are $\eta_{L1}=\eta_{R1}=0.5$, $\eta_{L2}=0.3$, $\eta_{R2}=0.9$, $\eta_{L21}=0.06$, $\eta_{L12}=0.04$, $\eta_{R21}=0.1$, $\eta_{R12}=0.3$, $\grave{o}_{21}=4$, $\grave{o}_{12}=3$ and $\mu_0=0$ all in units of meV.

In Fig. 5 (a) the comparison between the currents without coherency and with coherency (in approximate and general formula) for superconducting leads are depicted. This indicates the increase in currents due to the coherency and also more increasing according to the interdot coupling by general formula. The same interchange of interdot energy coupling ($\grave{o}_{21}=\grave{o}_{12}$) observes a pick by the approximate formula while the different interdot coupling remarks double picks in both $\grave{o}_{12}$ and $\grave{o}_{21}$ due to the general formula. Part (b) shows the effect of superconducting energy gap change in currents by general formula which expresses arising current level with the increase of energy gap (energy gap for normal state is zero).

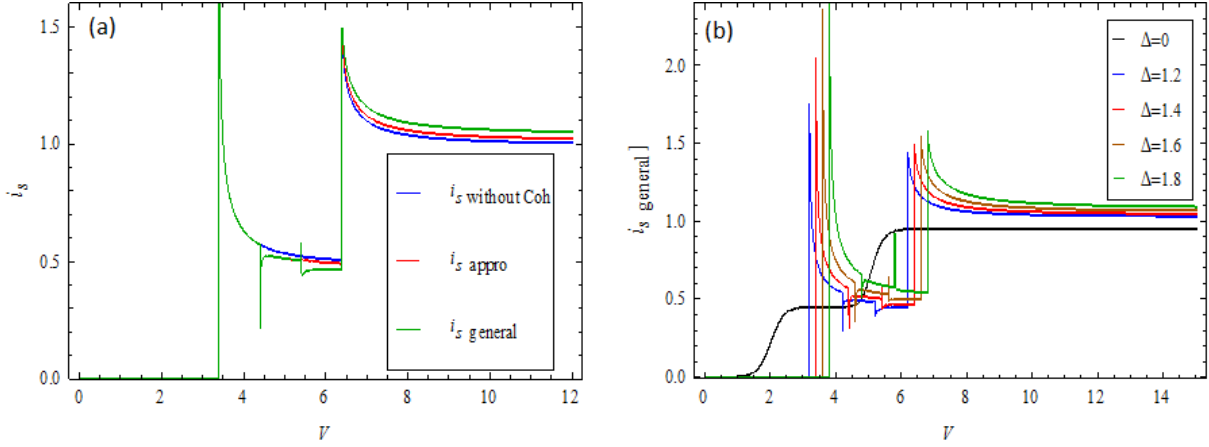

Fig.5.Steady state currents of superconducting leads. (a) The comparison with the steady state currents without coherences, approximate and general formula ($\Delta_L = \Delta_R = 1.4 meV$). (b) The variation of general formula current with the superconducting energy gap $\Delta$. The values of parameters in Fig. 4 are considered.

The time evolution of the populations is shown in Fig. 6 (a): solid line for normal and dashed line for superconducting leads. The dynamics of real and imaginary parts of coherency for normal leads are presented in Fig.6 (b) (Eq. (C.1)). It can be noted that the population and coherency are decoupled in the RWA. The populations evolve exponentially and reach a steady state distribution described by the eigenvector with zero eigenvalue but the coherences show a damped oscillatory behavior and vanish at long times.

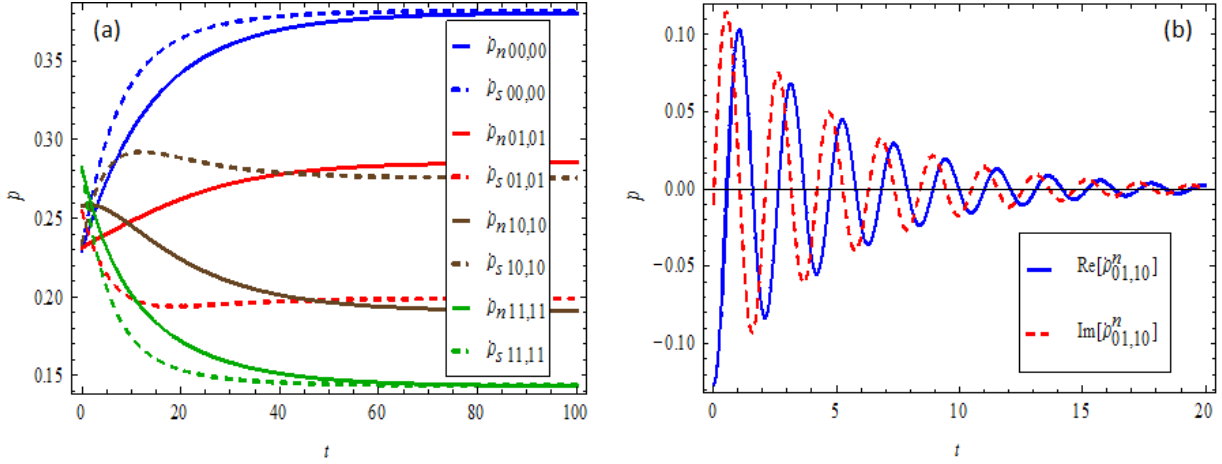

Fig.6.The population and coherencies dynamics. (a) Solid line and dashed line are the population of normal and superconducting leads respectively. (b) Solid line and dashed line are the real and imaginary part of coherency of normal leads respectively. The couplings are $\eta_{L1} = 0.01$, $\eta_{R1} = 0.02, \eta_{L2} = 0.03$, $\eta_{R2} = 0.04$, $\eta_{L21} = 0.06$, $\eta_{L12} = 0.04$, $\eta_{R21} = 0.1$, $\eta_{R12} = 0.3$, $\omega_{21} = 4$, $\omega_{12} = 3$, $V = 7$ and $\Delta = 1.5$ for part (a) for all in units of meV.

Fig. 7 describes the effect of energy gap in real part of coherency dynamics. It is obvious that the imaginary part has the same oscillatory behavior with only phase difference. We can conclude that the increase of energy gap extends the amplitude of coherency considerably and also makes the specific time of decay more long. It should be mentioned that the resonant energy of system affects the behavior of coherency with energy gap, which help us to find a proper energy gap to get the best results for coherency and current.

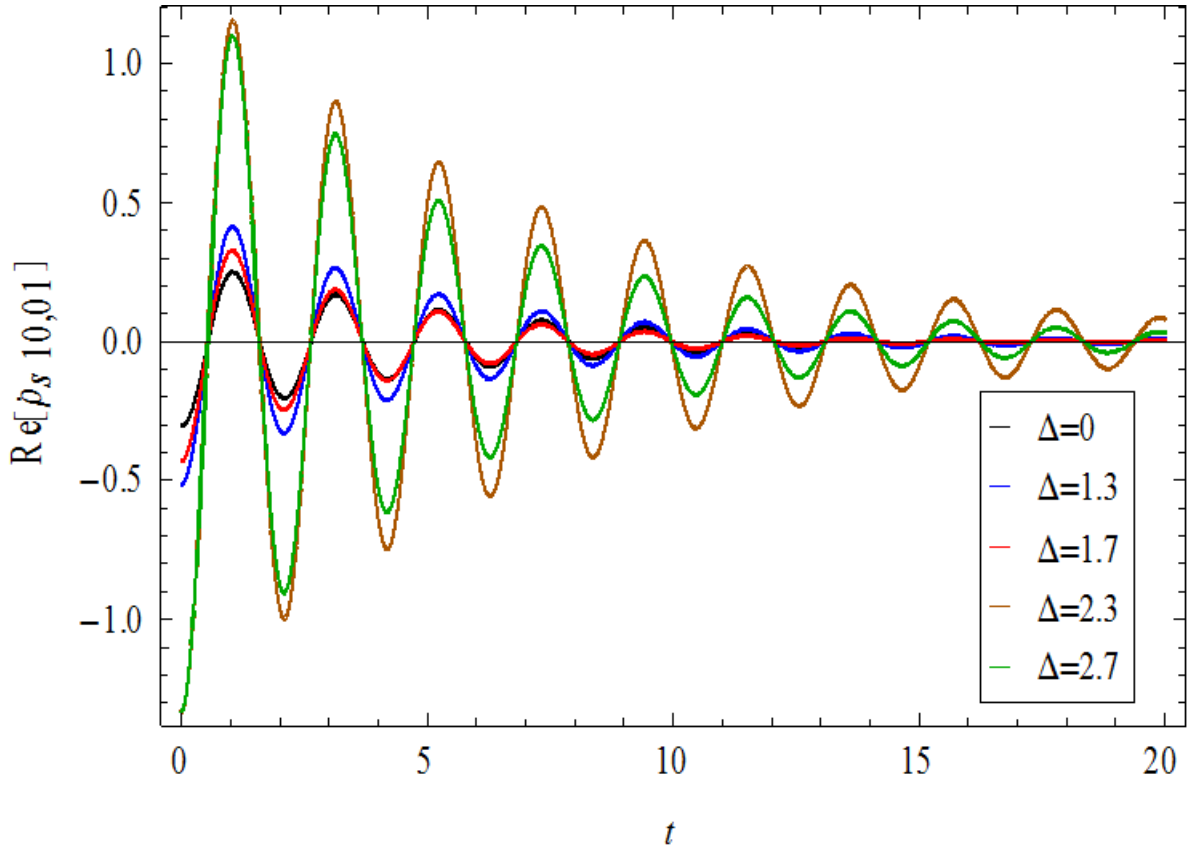

Fig.7.The superconducting energy gap variation with the real part of time dependent coherency for superconducting leads. The coupling parameters are the same as Fig. 6, and $V = 5$

5.  Conclusions
We have used a quantum master equation for the electron transport through two coupled QDs connected two superconducting reservoirs with different chemical potentials. We have shown that the reduced master equation has dissipator and the Lamb shift terms originating from the superconducting leads under the Born-Markov approximation. By ignoring and by considering the off-diagonal coherent elements of 6×6 $\hat{M}$ matrix we have obtained the nonequilibrium steady state density matrix, the electric current and coherencies through the system. We have also considered the inter-dots coupling energy which has the significant difference of double quantum dot to single one as shown in Fig. 4. This consideration have impressed by peaks in the currents and coherencies curves remarkably. In our investigation, the superconducting energy gap of reservoirs has affected the lifetime and amplitude of coherency oscillation by decreasing the dissipation. The energy gap increases the amplitude of coherency considerably and also makes the specific time of decay longer. Resonant energy of the system affects the behavior of coherency with energy gap, which help us to find a proper energy gap to get the best results for coherency and current

By increasing, the coherency and using the specific coupling energies of DQD (general formula) the electrical transport currents rise inevitably. Therefore, we can conclude that the energy gap provides the best results for coherency and current by fitting the physical parameters.

To extend the present model to non-Markovian quantum master equation, d-wave superconductors (HTSC) as leads, double parallel QDs or a chain of quantum dots as system and the thermoelectric current with different temperature of reservoirs are under our consideration and will be published elsewhere.

Acknowledgments
The authors wish to thank the Office of Graduate Studies and Research Vice President of the University of Isfahan for their support.

Appendix A. Derivation of quantum master equation

In the interaction picture we have $A_\alpha(\omega,t) = e^{-i\omega t} A_\alpha(\omega)$, $A_\alpha^\dagger(\omega,t) = e^{i\omega t} A_\alpha^\dagger(\omega)$, thus $A_\alpha^\dagger(\omega) = A_\alpha(-\omega)$. The summation over energy and using the completeness relation we get

$$\sum_\omega A_\alpha(\omega) = \sum_\omega A_\alpha^\dagger(-\omega) = A_\alpha \tag{A.1}$$

The interaction Hamiltonian becomes

$$H_I = \sum_{\alpha,\omega} A_\alpha(\omega) \otimes B_\alpha = \sum_{\alpha,\omega} A_\alpha^\dagger(-\omega) \otimes B_\alpha^\dagger \tag{A.2}$$

In general, the frequency spectrum $\{\omega\}$ degenerates so that for a fixed $\omega$ the index $\alpha$ labels the different operators $A_\alpha(\omega)$ belonging to the same frequency.

The Markovian master equation is improved from von Neumann in interaction picture. According the Markov approximation, the reservoir correlation functions decay sufficiently fast over the bath correlation time which is small compared to the relaxation time. So we can write

$$\frac{d\rho_s(t)}{dt} = \sum_\omega \sum_{\alpha,\beta} \Gamma_{\alpha\beta}(\omega) \left[ A_\beta(\omega) \rho_s(t) A_\alpha^\dagger(\omega) - A_\alpha^\dagger(\omega) A_\beta(\omega) \rho_s(t) \right] + h.c. \tag{A.3}$$

Here *h.c.* means the Hermitian conjugated expression. We have introduced the Fourier transforms as

$$\Gamma_{\alpha\beta}(\omega) = \int_0^\infty e^{i\omega s} ds \langle B_\alpha^\dagger(t) B_\beta(t-s) \rangle \tag{A.4}$$

In the stationary state of the reservoir $[H_B, \rho_B] = 0$, thus the reservoir correlation functions are homogeneous in time and we have

$$\langle B_\alpha^\dagger(t) B_\beta(t-s) \rangle = \langle B_\alpha^\dagger(s) B_\beta(0) \rangle \equiv tr_B \{ B_\alpha^\dagger(s) B_\beta(0) \rho_B \} \tag{A.5}$$

Appendix B. $\hat{M}$ matrix elements and preparing to calculate currents

Here $\hat{M}$ matrix elements of Eq. (22) are mentioned

$M_{11} = -2\gamma_{11}(-\dot{q}_1,-\dot{q}_1) - 2\gamma_{22}(-\dot{o}_2,-\dot{o}_2)$, $M_{12} = 2\gamma_{22}(\dot{o}_2,\dot{o}_2)$, $M_{13} = 2\gamma_{11}(\dot{q}_1,\dot{q}_1)$, $M_{14} = 0$, $M_{15} = 2\gamma_{21}(\dot{o}_2,\dot{q}_1)$, $M_{16} = 2\gamma_{12}(\dot{q}_1,\dot{o}_2)$, $M_{21} = 2\gamma_{22}(-\dot{o}_2,-\dot{o}_2)$, $M_{22} = -2\gamma_{11}(-\dot{q}_1,-\dot{q}_1) - 2\gamma_{22}(\dot{o}_2,\dot{o}_2)$, $M_{23} = 0$, $M_{24} = 2\gamma_{11}(\dot{q}_1,\dot{q}_1)$, $M_{25} = -\gamma_{12}(\dot{q}_1,\dot{o}_2) - \gamma_{21}(-\dot{o}_2,-\dot{q}_1) - i\dot{q}_{12}$, $M_{26} = -\gamma_{12}(-\dot{q}_1,-\dot{o}_2) - \gamma_{21}(\dot{o}_2,\dot{q}_1) + i\dot{q}_{21}$, $M_{31} = 2\gamma_{11}(-\dot{q}_1,-\dot{q}_1)$, $M_{32} = 0$, $M_{33} = -2\gamma_{11}(\dot{q}_1,\dot{q}_1) - 2\gamma_{22}(-\dot{o}_2,-\dot{o}_2)$, $M_{34} = 2\gamma_{22}(\dot{o}_2,\dot{o}_2)$, $M_{35} = -\gamma_{12}(\dot{q}_1,\dot{o}_2) - \gamma_{21}(-\dot{o}_2,-\dot{q}_1) + i\dot{q}_{12}$, $M_{36} = -\gamma_{12}(-\dot{q}_1,-\dot{o}_2) - \gamma_{21}(\dot{o}_2,\dot{q}_1) - i\dot{q}_{21}$, $M_{41} = 0$, $M_{42} = 2\gamma_{11}(-\dot{q}_1,-\dot{q}_1)$, $M_{43} = 2\gamma_{22}(-\dot{o}_2,-\dot{o}_2)$, $M_{44} = -2\gamma_{11}(\dot{q}_1,\dot{q}_1) - 2\gamma_{22}(\dot{o}_2,\dot{o}_2)$, $M_{45} = 2\gamma_{12}(-\dot{q}_1,-\dot{o}_2)$, $M_{46} = 2\gamma_{21}(-\dot{o}_2,-\dot{q}_1)$, $M_{51} = 2\gamma_{21}(-\dot{o}_2,-\dot{q}_1)$, $M_{52} = -\gamma_{12}(-\dot{q}_1,-\dot{o}_2) - \gamma_{21}(\dot{o}_2,\dot{q}_1) - i\dot{q}_{21}$,

$$M_{53} = -\gamma_{12}(-\dot{\omega}_1, -\dot{\omega}_2) - \gamma_{21}(\dot{\omega}_2, \dot{\omega}_1) + i\dot{\omega}_{21},  \qquad M_{54} = 2\gamma_{12}(\dot{\omega}_1, \dot{\omega}_2),$$

$$M_{55} = -\gamma_{11}^R(\dot{\omega}_1, \dot{\omega}_1) - \gamma_{11}^L(-\dot{\omega}_1, -\dot{\omega}_1) - \gamma_{22}^R(\dot{\omega}_2, \dot{\omega}_2) - \gamma_{22}^L(-\dot{\omega}_2, -\dot{\omega}_2) + i(\dot{\omega}_{22} - \dot{\omega}_{11}), \quad M_{56} = 0, \ M_{61} = 2\gamma_{12}(-\dot{\omega}_1, -\dot{\omega}_2),$$

$$M_{62} = -\gamma_{12}(\dot{\omega}_1, \dot{\omega}_2) - \gamma_{21}(-\dot{\omega}_2, -\dot{\omega}_1) + i\dot{\omega}_{12}, \qquad M_{63} = -\gamma_{12}(\dot{\omega}_1, \dot{\omega}_2) - \gamma_{21}(-\dot{\omega}_2, -\dot{\omega}_1) - i\dot{\omega}_{12}, \qquad M_{64} = 2\gamma_{21}(\dot{\omega}_2, \dot{\omega}_1),$$

$$M_{65} = 0, \ M_{66} = -\gamma_{11}^R(\dot{\omega}_1, \dot{\omega}_1) - \gamma_{11}^L(-\dot{\omega}_1, -\dot{\omega}_1) - \gamma_{22}^R(\dot{\omega}_2, \dot{\omega}_2) - \gamma_{22}^L(-\dot{\omega}_2, -\dot{\omega}_2) - i(\dot{\omega}_{22} - \dot{\omega}_{11}).$$

To calculate the currents with coherency in general form (Eqs. (25) and (26)), we use $N = \begin{pmatrix} 0 & 1 & 1 & 2 & 0 & 0 \end{pmatrix}$, $M_L^{in}[\gamma^L(-\dot{\omega})]$ and $M_L^{out}[\gamma^L(\dot{\omega})]$ matrices (utilizing the $\hat{M}$ matrix)

Appendix C. Population and coherencies time evolution in RWA

In the RWA, populations and coherences evolve independently. The population dynamics and the time-dependence of coherences are determined as

$$\rho_{00,00}(t) = g_1 \frac{\gamma_{11}^R(-\dot{\omega}_1, -\dot{\omega}_1)\gamma_{22}^R(-\dot{\omega}_2, -\dot{\omega}_2)}{\gamma_{11}^L(-\dot{\omega}_1, -\dot{\omega}_1)\gamma_{22}^L(-\dot{\omega}_2, -\dot{\omega}_2)} e^{\varepsilon_1 t} - g_2 \frac{\gamma_{22}^R(-\dot{\omega}_2, -\dot{\omega}_2)}{\gamma_{22}^L(-\dot{\omega}_2, -\dot{\omega}_2)} e^{\varepsilon_2 t} - g_3 \frac{\gamma_{11}^R(-\dot{\omega}_1, -\dot{\omega}_1)}{\gamma_{11}^L(-\dot{\omega}_1, -\dot{\omega}_1)} e^{\varepsilon_3 t} + g_4 e^{\varepsilon_4 t}$$

$$\rho_{01,01}(t) = g_1 \frac{\gamma_{11}^R(-\dot{\omega}_1, -\dot{\omega}_1)}{\gamma_{11}^L(-\dot{\omega}_1, -\dot{\omega}_1)} e^{\varepsilon_1 t} - g_2 e^{\varepsilon_2 t} + g_3 \frac{\gamma_{11}^R(-\dot{\omega}_1, -\dot{\omega}_1)}{\gamma_{11}^L(-\dot{\omega}_1, -\dot{\omega}_1)} e^{\varepsilon_3 t} - g_4 e^{\varepsilon_4 t}$$

$$\rho_{10,10}(t) = g_1 \frac{\gamma_{22}^R(-\dot{\omega}_2, -\dot{\omega}_2)}{\gamma_{22}^L(-\dot{\omega}_2, -\dot{\omega}_2)} e^{\varepsilon_1 t} + g_2 \frac{\gamma_{22}^R(-\dot{\omega}_2, -\dot{\omega}_2)}{\gamma_{22}^L(-\dot{\omega}_2, -\dot{\omega}_2)} e^{\varepsilon_2 t} - g_3 e^{\varepsilon_3 t} - g_4 e^{\varepsilon_4 t} \qquad (C.1)$$

$$\rho_{11,11}(t) = g_1 e^{\varepsilon_1 t} + g_2 e^{\varepsilon_2 t} + g_3 e^{\varepsilon_3 t} + g_4 e^{\varepsilon_4 t}$$

$$\rho_{10,01}(t) = g_5 e^{\varepsilon_5 t}$$

$$\rho_{01,10}(t) = g_6 e^{\varepsilon_6 t}$$

where the coefficients $g_1$-$g_4$ are related to the initial density matrix as follows

$$g_1 = \frac{\gamma_{11}^R(-\dot{\omega}_1, -\dot{\omega}_1)\gamma_{22}^R(-\dot{\omega}_2, -\dot{\omega}_2)}{\gamma_{11}^L(-\dot{\omega}_1, -\dot{\omega}_1)\gamma_{22}^L(-\dot{\omega}_2, -\dot{\omega}_2)} \rho_{00,00}(0) + \frac{\gamma_{11}^R(-\dot{\omega}_1, -\dot{\omega}_1)}{\gamma_{11}^L(-\dot{\omega}_1, -\dot{\omega}_1)} \rho_{01,01}(0) + \frac{\gamma_{22}^R(-\dot{\omega}_2, -\dot{\omega}_2)}{\gamma_{22}^L(-\dot{\omega}_2, -\dot{\omega}_2)} \rho_{10,10}(t) + \rho_{11,11}(t)$$

$$g_2 = -\frac{\gamma_{22}^R(-\dot{\omega}_2, -\dot{\omega}_2)}{\gamma_{22}^L(-\dot{\omega}_2, -\dot{\omega}_2)} \rho_{00,00}(0) - \rho_{01,01}(0) + \frac{\gamma_{22}^R(-\dot{\omega}_2, -\dot{\omega}_2)}{\gamma_{22}^L(-\dot{\omega}_2, -\dot{\omega}_2)} \rho_{10,10}(t) + \rho_{11,11}(t)$$

$$g_3 = -\frac{\gamma_{11}^R(-\dot{\omega}_1, -\dot{\omega}_1)}{\gamma_{11}^L(-\dot{\omega}_1, -\dot{\omega}_1)} \rho_{00,00}(0) + \frac{\gamma_{11}^R(-\dot{\omega}_1, -\dot{\omega}_1)}{\gamma_{11}^L(-\dot{\omega}_1, -\dot{\omega}_1)} \rho_{01,01}(0) - \rho_{10,10}(t) + \rho_{11,11}(t) \qquad (C.2)$$

$$g_4 = \rho_{00,00}(0) - \rho_{01,01}(0) - \rho_{10,10}(t) + \rho_{11,11}(t)$$

$$g_5 = \rho_{10,01}(0)$$

$$g_6 = \rho_{01,10}(0)$$

where $\rho_{00,00}(0) + \rho_{01,01}(0) + \rho_{10,10}(0) + \rho_{11,11}(0) = 1$ and $\rho_{10,01} = \rho_{01,10}^*$.

Corresponding author Tel.: +98 3117934732; fax: +98 3117922409.
E-mail addresses: h.yavary@sci.ui.ac.ir